\documentclass{article}

\usepackage[dblblindworkshop, preprint]{neurips_2026}

\usepackage[utf8]{inputenc} 
\usepackage[T1]{fontenc}    
\usepackage{hyperref}       
\usepackage{url}            
\usepackage{booktabs}       
\usepackage{amsfonts}       
\usepackage{nicefrac}       
\usepackage{microtype}      
\usepackage{xcolor}         
\usepackage{graphicx}

\title{Adversarial Entropy Inflation Against Gumbel-Based Inference Verification}
\workshoptitle{NeurIPS 2026 Workshop FLMSec}

\author{%
  Nikita Kezins \\
  Delft University of Technology\\
  \texttt{nikitakezins@gmail.com}
}

\begin{document}

\maketitle

\begin{abstract}
Gumbel-based inference verification bounds LLM weight exfiltration by only forgiving token choices that plausibly arise from honest GPU nondeterminism, reporting a $>200\times$ slowdown for a steganographic adversary under benign prompt traffic. This bound assumes a passive attacker; we show it degrades sharply against an adversary who instead controls the prompt distribution. Because the verifier's admissible-token-set size is driven by the model's own output entropy, prompts engineered to break grammatical and sub-word structure -- rather than benign conversational traffic -- widen that set and open a materially larger covert channel. Across six instruction-tuned models spanning 1B to 32B parameters and three random seeds, our strongest attack (character- and script-level disruption) roughly doubles bits leaked per token relative to benign prompts, cutting the slowdown factor to $60$--$118\times$. These results indicate that static, benign-traffic-calibrated thresholds are insufficient for this defense, and that jitter-forgiveness thresholds should instead be calibrated dynamically against local token entropy.
\end{abstract}

\section{Introduction}
As organizations scale their artificial intelligence capabilities, proprietary models become highly valuable assets, raising significant concerns regarding model weight exfiltration. One vector for such exfiltration is through standard API inference, where malicious users can encode and leak model weights via deliberate token selection. This technique effectively establishes a covert subliminal channel ~\cite{simmons1984prisoners} within the model's standard output stream.

As a defense against this, ~\cite{rinberg2026verifyingllminferencedetect} proposed verifying sampled tokens under a predefined seed selection, building on the same Gumbel-margin statistic used more broadly for detecting inference misconfiguration and tampering~\cite{karvonen2025difrinferenceverificationdespite}. By utilizing Gumbel sampling, their method ensures nearly identical token selection under an argmax policy for both the verifier and the prover. While theoretically sound, this approach is vulnerable to GPU non-determinism (jitter) caused by non-associative floating-point operations. When multiple tokens have near-identical likelihoods, this jitter can flip the argmax token selection. The necessity for verifiers to forgive these jitter-induced discrepancies inadvertently creates an attack gap that an active exfiltrator can exploit.

Prior work demonstrated that under benign input conditions (e.g., a subset of the LMSYS Chatbot Arena dataset~\cite{zheng2023lmsyschat1m}), only a small fraction of tokens exhibit branching -- where multiple tokens share similar likelihoods, allowing GPU jitter to alter the output -- with the rest exactly determined (see Methodology for the formal admissible-set definition). This underlies the reported 200x slowdown in exfiltration rates relative to the theoretical maximum.

Because the capacity of this covert channel is strictly bound by the entropy of the token distribution, widening the attack gap requires an exfiltrator to maximize output randomness. In this work, we demonstrate that by deliberately breaking token dependencies and maximizing output entropy via adversarial prompting, an attacker can double the exfiltration capacity, reducing the defense's slowdown factor from 200x to approximately 100x.

\section{Methodology}

\textbf{Preliminaries.} The GLS verifier of \cite{rinberg2026verifyingllminferencedetect} reconstructs, per output position, the Gumbel-perturbed logits the prover's sampler would use under a shared seed, then defines the \emph{admissible set} $A_t$ as tokens within tolerance $\sigma$ of the top Gumbel score -- i.e.\ tokens a compliant prover could plausibly emit under GPU nondeterminism. Verification passes iff the observed token lies in $A_t$; residual exfiltration bandwidth at that position is bounded by $\log_2 |A_t|$ bits. On benign traffic $A_t$ is a singleton (``forced'') almost everywhere, giving the reported $>200\times$ slowdown. Since $|A_t|$ tracks the model's own output entropy, an adversary who raises that entropy through prompt design directly widens $A_t$ -- the attack surface we study.

We build upon the experimental framework provided by \cite{rinberg2026verifyingllminferencedetect}. To establish a baseline, we evaluate a subset of 150 benign prompts from the LMSYS dataset~\cite{zheng2023lmsyschat1m}, matching the original study. We then design three tiers of adversarial attacks, each progressively escalating the attempt to induce high-entropy distributions. We generate 150 adversarial prompts for each attack type.

\textbf{TERSE (Low-Context Generation):} 722 combinatorially generated one-word fill-in-the-blank and low-context prompts (e.g.\ single-word mood/preference/description completions, terse conversational openers). The hypothesis is that genuine uncertainty concentrates on the first token of a response; isolating it rather than diluting it across a longer sequence should widen the set of admissible tokens.

\textbf{SCRAMBLE (Grammatical Dependency Disruption):} 233 prompts across four categories: unrelated-item lists with explicit anti-correlation instructions, pseudo-random password/hash/PIN generation, gibberish-syllable continuation, and short surrealist continuations. This targets local word-to-word grammatical dependencies, fundamentally undermining the structure that makes fluent prose low-entropy.

\textbf{CIPHER (Sub-Word and Token-Level Disruption):} 157 prompts across four categories: unbroken alphanumeric blocks, base64/hex/base32 dumps, gibberish-prefix ``context bombs'' the model must continue verbatim, and multi-script character collisions (Latin/Cyrillic/Katakana/Arabic). Because even ``unrelated'' word lists rely on standard vocabulary, this attack targets character-level distributions directly to maximize entropy at the token boundary.

We executed these four prompt sets (benign, plus three attacks; 150 prompts each) across three distinct random seeds. To ensure generalizability across different architectures and scales, we evaluated six instruction-tuned models: Llama-3.2-1B, Llama-3.2-3B, Llama-3.1-8B, and Qwen2.5-14B, Qwen3-30B-MoE, Qwen3-32B.

\section{Experiments and Results}

\begin{figure}[ht]
    \centering
    \begin{minipage}{0.48\linewidth}
        \centering
        \includegraphics[width=\linewidth]{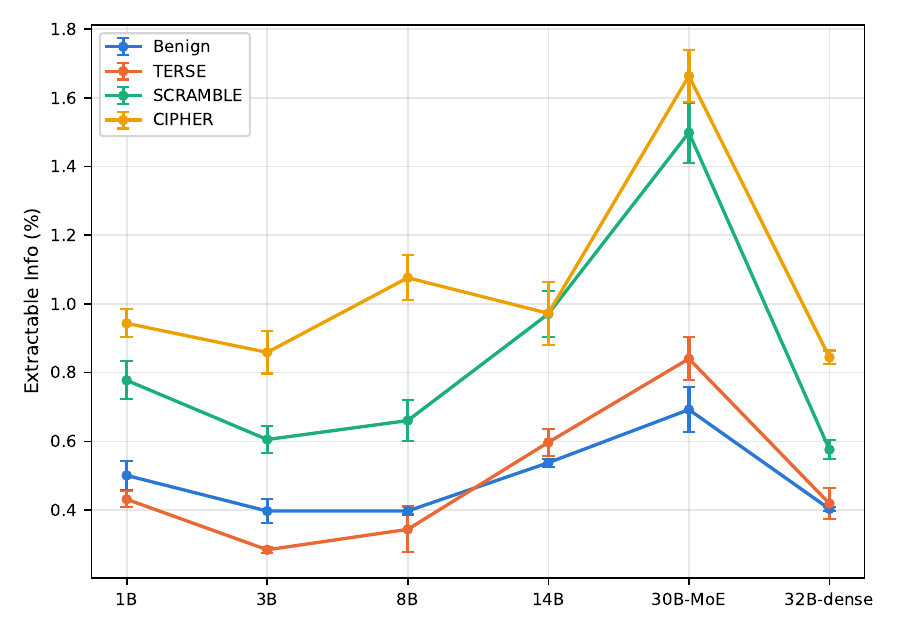}
    \end{minipage}\hfill
    \begin{minipage}{0.48\linewidth}
        \centering
        \includegraphics[width=\linewidth]{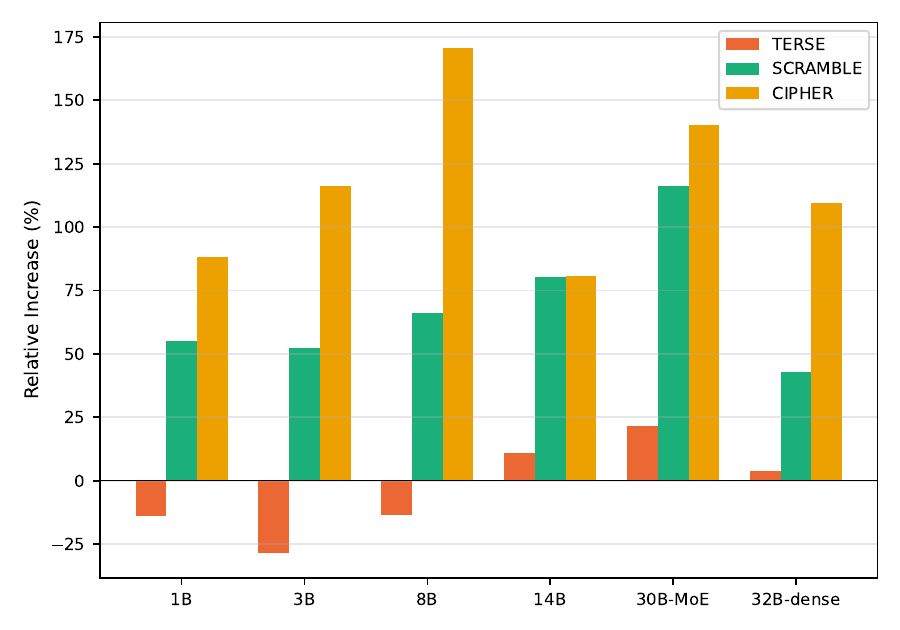}
    \end{minipage}
    
    \vspace{0.1cm}

    \includegraphics[width=0.82\linewidth]{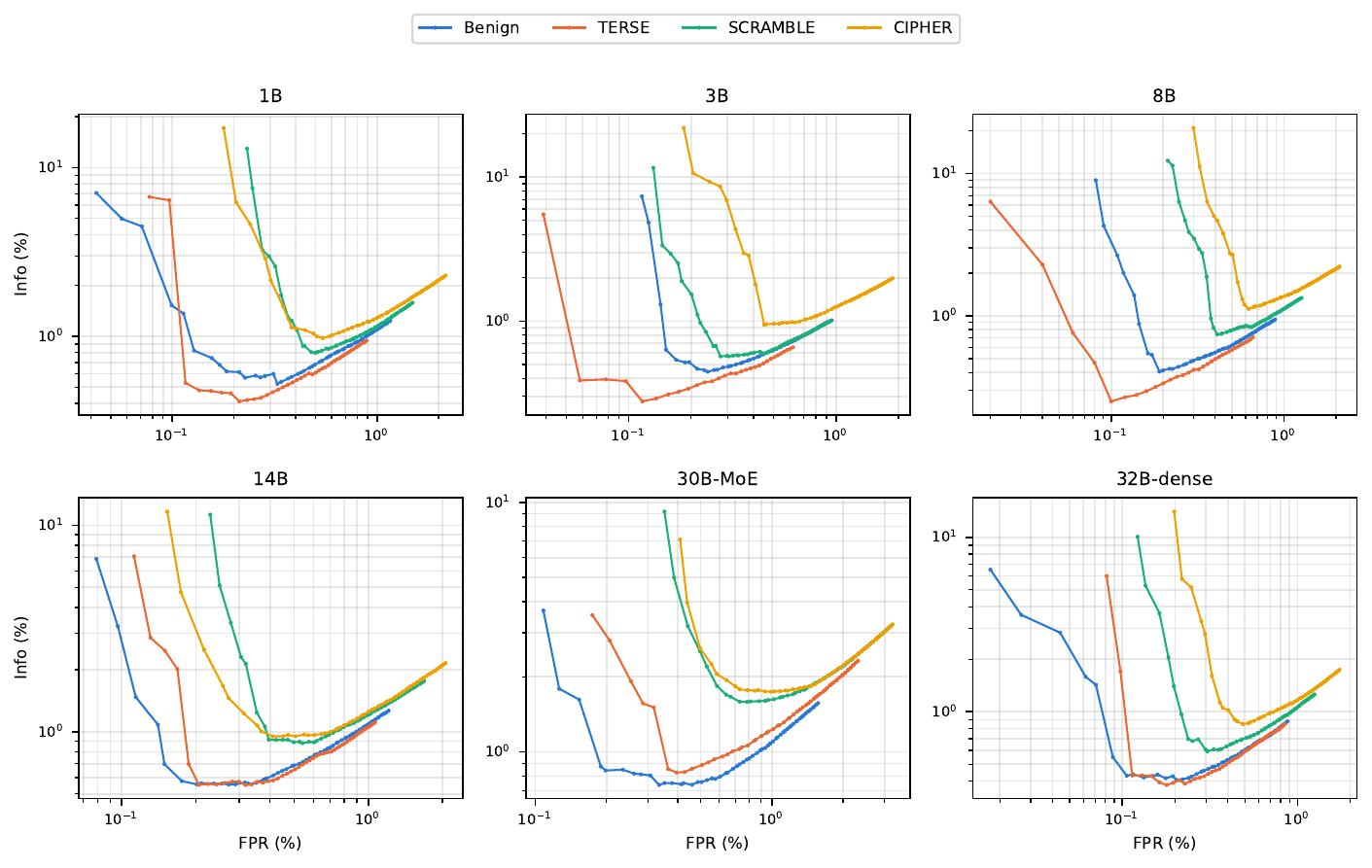}
    \vspace{-0.35cm}
    \caption{\textbf{Top Left:} Extractable info (vocab-size-corrected) at verifier tolerance $\sigma=1.0$. \textbf{Top Right:} Relative exfiltration gain vs. benign baseline. \textbf{Bottom:} FPR vs. Extractable Info Pareto curves. CIPHER maximizes covert bandwidth; TERSE collapses on smaller models.}
    \label{fig:combined_results}
\end{figure}

\begin{table}[ht]
  \caption{Impact of adversarial prompts on exfiltration bandwidth and defense slowdown. CIPHER significantly increases bits extracted per token and drastically reduces the verifier's slowdown multiplier compared to the benign baseline.}
  \label{tab:main_results}
  \centering
  \resizebox{\textwidth}{!}{
  \renewcommand{\arraystretch}{0.82}
  \begin{tabular}{lcccc|cccc}
    \toprule
    & \multicolumn{4}{c}{\textbf{Bits Extracted per Token} ($\uparrow$)} & \multicolumn{4}{c}{\textbf{Slowdown Multiplier} ($\downarrow$)} \\
    \cmidrule(lr){2-5} \cmidrule(lr){6-9}
    \textbf{Model} & Benign & TERSE & SCRAMBLE & CIPHER & Benign & TERSE & SCRAMBLE & CIPHER \\
    \midrule
    1B & 0.085$\pm$0.007 & 0.073$\pm$0.004 & 0.132$\pm$0.009 & 0.160$\pm$0.007 & 201$\pm$17$\times$ & 232$\pm$12$\times$ & 129$\pm$9$\times$ & 106$\pm$5$\times$ \\
    3B & 0.067$\pm$0.006 & 0.048$\pm$0.002 & 0.103$\pm$0.007 & 0.146$\pm$0.011 & 254$\pm$22$\times$ & 352$\pm$10$\times$ & 166$\pm$11$\times$ & 117$\pm$8$\times$ \\
    8B & 0.068$\pm$0.002 & 0.058$\pm$0.011 & 0.112$\pm$0.010 & 0.183$\pm$0.011 & 252$\pm$7$\times$  & 304$\pm$68$\times$ & 153$\pm$14$\times$ & 93$\pm$6$\times$ \\
    14B & 0.093$\pm$0.002 & 0.103$\pm$0.007 & 0.167$\pm$0.012 & 0.167$\pm$0.016 & 186$\pm$4$\times$  & 168$\pm$11$\times$ & 103$\pm$7$\times$ & 104$\pm$9$\times$ \\
    30B-MoE & 0.119$\pm$0.011 & 0.145$\pm$0.011 & 0.258$\pm$0.015 & 0.286$\pm$0.013 & 146$\pm$15$\times$ & 120$\pm$9$\times$  & 67$\pm$4$\times$  & \textbf{60$\pm$3$\times$} \\
    32B-dense & 0.070$\pm$0.001 & 0.072$\pm$0.008 & 0.099$\pm$0.005 & 0.145$\pm$0.003 & 248$\pm$4$\times$  & 241$\pm$24$\times$ & 174$\pm$8$\times$ & \textbf{118$\pm$3$\times$} \\
    \bottomrule
  \end{tabular}
  }
\end{table}

As shown in Figure~\ref{fig:combined_results} and Table~\ref{tab:main_results}, adversarial prompt design significantly impacts the efficacy of Gumbel-sampling-based verification. We highlight three core findings:

\textbf{1. Shattering the Slowdown Multiplier:} Table~\ref{tab:main_results} validates the claimed $>200\times$ benign-case slowdown (e.g., $248\times$ on 32B-dense), but CIPHER systematically dismantles it, dropping the slowdown to $118\times$ on 32B-dense and $60\times$ on 30B-MoE.

\textbf{2. Sub-word Disruption Maximizes Leakage:} CIPHER (context-bomb/multi-script) consistently achieves the highest exfiltration capacity. By attacking character-level priors, it prevents models from collapsing onto common vocabulary, yielding up to a 170\% relative increase over benign prompts and peaking at 0.286 bits extracted per token on Qwen3-30B-MoE.

\textbf{3. Low-Context Prompts Induce Collapse:} TERSE underperforms the benign baseline on sub-14B models. Restricting context causes smaller models to default to high-probability tokens, ironically collapsing the output distribution into a near-deterministic state and exacerbating the defense's slowdown (peaking at $352\times$ on the 3B model).

\section{Conclusion}
Gumbel-based verification's security guarantees are heavily contingent on output entropy: adversarial prompts that disrupt syntactic and sub-word priors widen the jitter-induced attack gap, slashing the purported $200\times$ slowdown to $60\times$--$118\times$ depending on architecture. Verifier designs cannot rely on static, benign-calibrated baselines; jitter forgiveness thresholds must instead be calibrated dynamically against local token entropy. This is not unique to the unilateral weight-exfiltration setting -- broader bilateral compute-verification architectures built on the same unexplained-information bound~\cite{petrie2026verifying} inherit an identical entropy-dependent attack surface.

\begin{ack}
This work was conducted during the Hardware Assurance Program at the Cambridge AI Safety Hub (CAISH).
\end{ack}

\bibliographystyle{plainnat}
\bibliography{references}

\appendix

\end{document}